\documentstyle[bibnorm]{lamuphys}
\newcommand{\be}{\begin{equation}}
\newcommand{\qee}{\end{equation}}
\newenvironment{emphit}{\begin{itemize}\em}{\end{itemize}}
\begin{document}

\title{The QCD  string  and generalized  wave equation}

\author{George K.\,Savvidy}

\institute{National Research Center Demokritos, Agia Paraskevi, GR-15310 Athens, Greece}

\maketitle

\begin{abstract}
The equation for QCD string proposed earlier is reviewed. This 
equation appears when we examine the gonihedric string model and the 
corresponding transfer matrix. Arguing that string equation should 
have a generalized Dirac form we found the corresponding infinite-dimensional 
gamma matrices as a symmetric solution of the Majorana commutation relations. 
The generalized gamma matrices are anticommuting and guarantee unitarity 
of the theory at all orders of $v/c$. In the second quantized form the equation 
does not have unwanted ghost states in Fock space. In the absence of Casimir mass 
terms the spectrum reminds hydrogen exitations. On every mass level $r=2,4,..$
there are different charged particles with spin running from $j=1/2$ up to 
$j_{max}=r-1/2$, and the degeneracy is equal to $d_{r}=2r-1 = 2j_{max}$.
This is in contrast with the exponential degeneracy in superstring theory.
\end{abstract}

\subsubsection{Introduction.}There is some experimental and theoretical evidence 
for the existence of a
string theory in four dimensions which may describe strong interactions and 
represent the solution of QCD \cite{nilsen}. 

One of the possible candidates for that
purpose is the gonihedric string which has been defined as a model of 
random surfaces with an action which is proportional to the linear size 
of the surface \cite{sav2}
\be
A(M) = m\sum_{<ij>} \lambda_{ij}
\cdot \Theta(\alpha_{ij}),~~~~~~~ \Theta(\alpha)= \vert 
\pi - \alpha \vert^{\varsigma} , \label{action}
\qee
where $\lambda_{ij}$ is the length of the edge $<ij>$ of the 
triangulated surface $M$ and $\alpha_{ij}$ is the dihedral angle 
between two neighbouring triangles of $M$ sharing a common edge $<ij>$.
The model has a number of properties which make it very close to the Feynman
path integral for a point-like relativistic particle. 
In the limit in which the surface degenerates into a single 
world line the action becomes proportional to the length of the 
path and the classical equation of motion for the gonihedric string
is reduced to the classical equation of motion for a free relativistic particle. 
At the classical level the string tension is equal to zero and, 
as it was demonstrated in \cite{sav2}, quantum fluctuations generate the 
nonzero string tension 
$
\sigma_{quantum}= \frac{d}{a^2}~(1 - ln \frac{d}{\beta}) , 
$
where $d$ is the dimension of the spacetime, $\beta$ is the coupling constant,
$a$ is the scaling parameter and $\varsigma = (d-2)/d$ in 
(\ref{action}).

It is natural therefore to ask what type of equation may describe this string
theory in the continuum limit.
The aim of the articles \cite{sav13,sav14} was to suggest a possible answer to this
question. The analysis of the 
transfer matrix shows \cite{sav13} that the desired equation 
should describe propagation of fermionic degrees of freedom distributed 
over the space contour. When this contour shrinks to a point, the equation 
should describe propagation of a free Dirac fermion. Thus each
particle in this theory should be viewed as a state of a 
complex fermionic system and the system should have a point-particle 
limit when there is no excitation of the internal motion. In the given case 
this restriction should be 
understood as a principle according to which  the infinite sequence of 
particles should contain the spin one-half fermion and the 
equation should have the Dirac form  \cite{sav13,sav14}

\be
\{~i~\Gamma_{\mu}~\partial^{\mu}~~-~~M~\}~~\Psi~~~=0 . \label{stringeq}
\qee
The invariance of this equation under Lorentz transformations
$x'_{\mu} =\Lambda_{\mu}^{~\nu}~ x_{\nu},~~~\Psi'(x')=
\Theta(\Lambda)~ \Psi(x)$
leads to the equation for the gamma matrices \cite{dirac,majorana}
$
\Gamma_{\nu} = \Lambda_{\nu}^{~\mu}~\Theta~\Gamma_{\mu}~\Theta^{-1}.
$
If we use the infinitesimal form of Lorentz transformations
$\Lambda_{\mu \nu}= \eta_{\mu \nu} + \epsilon_{\mu \nu},~~~\Theta =
1 + \frac{1}{2}\epsilon_{\mu \nu}~I^{\mu \nu}$
it follows that gamma matrices should satisfy the Majorana commutation relation
\cite{majorana}
\be
[\Gamma_{\mu},~I_{\lambda \rho}] = \eta_{\mu \lambda}~\Gamma_{\rho}
- \eta_{\mu \rho}~\Gamma_{\lambda},  \label{mcr}
\qee
where $I_{\mu \nu}$ are the generators of the Lorentz algebra. These equations 
allow us to find the $\Gamma_{\mu}$ matrices when the representation of the 
$I_{\mu \nu}$ is given. 

Ettore Majorana in 1932 \cite{majorana} found a solution of the 
equations (\ref{mcr}) which differs from the Dirac 
gamma matrices \cite{dirac}. The original Majorana solution 
for $\Gamma_{\mu}$ matrices is infinite-dimensional (see equation (14) 
in \cite{majorana}) and is given by the formulas:
\begin{eqnarray}
<j,m\vert ~\Gamma_{0} ~\vert j,m> = j+1/2 \nonumber\\
<j,m\vert ~\Gamma_{z} ~\vert j+1,m> = \frac{i}{2}\sqrt{(j+m+1)(j-m+1)} \nonumber\\
<j,m\vert ~\Gamma_{z} ~\vert j-1,m> = -\frac{i}{2}\sqrt{(j+m)(j-m)} \nonumber\\
<j,m\vert ~\Gamma_{+} ~\vert j+1,m-1> = \frac{i}{2}\sqrt{(j-m+1)(j-m+2)} \nonumber\\
<j,m\vert ~\Gamma_{+} ~\vert j-1,m-1> = \frac{i}{2}\sqrt{(j+m)(j+m-1)} \nonumber\\
<j,m\vert ~\Gamma_{-} ~\vert j+1,m+1> = -\frac{i}{2}\sqrt{(j+m+1)(j+m+2)} \nonumber\\
<j,m\vert ~\Gamma_{-} ~\vert j-1,m-1> = -\frac{i}{2}\sqrt{(j-m)(j-m-1)} .\nonumber
\end{eqnarray}
One can see from this solution that the mass spectrum of the theory is 
equal to
\be
M_{j}=\frac{M}{j+1/2},               \label{majorana}
\qee
where $j=1/2,3/2,5/2,....$ in the fermion case and $j=0,1,2,....$ in the 
boson case.
The main problems of Majorana theory are the decreasing mass spectrum 
(\ref{majorana}), 
absence of antiparticles and troublesome tachyonic solutions - the problems 
common to high spin theories \cite{nambu,grodsky}.

An alternative way to incorporate the internal motion 
into the Dirac equation was suggested by Pierre Ramond in 1971 \cite{ramond}. 
In his extension of the Dirac equation the internal motion 
is incorporated through the construction of operator-valued gamma matrices. 
The equations which define the $\Gamma_{\mu}$ matrices are
$$
<\Gamma_{\mu}(\tau)> = \gamma_{\mu},~~~~~\{ \Gamma_{\mu}(\tau), 
\Gamma_{\nu}(\tau^{'}) \} = 
2 \eta_{\mu \nu} ~\delta(~ \frac{1}{2\pi\alpha^{'} } ~(\tau -\tau^{'})~ ),
$$
$
\Gamma^{+}_{\mu}(\tau)~\gamma_{0}=
\gamma_{0}~\Gamma_{\mu}(\tau),
$
where it is required that the proper-time average $<...>$ 
over the $periodic$ internal motion with period 
$2\pi \alpha^{'} =1 /\sigma $ should coincide with the Dirac 
matrices. The solution has the form 
$$
\Gamma_{\mu}(\tau) = \gamma_{\mu} + i\gamma_{5} 
\sum_{n=1}^{\infty}[b^{n}_{\mu}exp(-i n\omega\tau) +
b^{+n}_{\mu} exp(i n\sigma\tau)], 
$$
where $b$'s are operators obeying the anticommutation relations
$\{ b^{n}_{\mu},b^{+m}_{\nu}\}= 2g_{\mu\nu}\delta_{nm}$ and all 
others are set equal to zero ($\omega = 1/\alpha^{'}$).
The mass spectrum lies on linear trajectories and 
the free Ramond string is a consistent theory 
in ten dimensions and the spectrum contains a massless ground state
\cite{15years,green}.

In both cases one can see an effective extension of Dirac gamma
matrices into the infinite-dimensional case, but these extensions are 
quiet different \cite{sav13,sav14}. 
For our purposes we shall follow Majorana's approach to incorporate the 
internal motion in the form of an infinite-dimensional wave equation.
Indeed in \cite{sav13,sav14}   
the Majorana theory was interpreted as a natural way to incorporate 
additional degrees of freedom into the relativistic Dirac equation.   
Unlike Majorana the authors consider the infinite sequence of 
high-dimensional representations of the Lorentz group with  nonzero Casimir
operators $(\vec{a}\cdot\vec{b})$ and $(\vec{a}^2 -\vec{b}^2)$. 
These representations $(j_{0},\lambda)$ and 
their adjoint $(j_{0},-\lambda)$ are enumerated by the 
index $r=j_{0} +1/2$, where $r=1,...,N$ and $j_{0}=1/2,3/2,...$~ is 
the lower spin in the representation $(j_{0},\lambda)$, thus 
$j=j_{0},j_{0}+1,...$. We took the free complex parameter $\lambda$ in
the real interval $-3/2 \leq \lambda \leq 3/2 $ in order to have $real$
matrix elements for the Lorentz boost operator $\vec{b}$.  
These representations are infinite-dimensional
except of the case $j_{0}=1/2,~\lambda = \pm 3/2$. 

The {\it duality transformation } 
$(j_{0};\lambda)  \rightarrow (\lambda;j_{0})$, defined in \cite{sav13}, 
leads to a $subsequent$ $restriction$ on the free parameter $\lambda$ and requires 
$\lambda =1/2$, so that the dual representations $\Theta_{r},\Theta_{\dot{r}}$ 
become finite-dimensional $(1/2, \pm (1/2 + r))$. The corresponding 
equation found in \cite{sav13} is not in contradiction with no-go theorem of
\cite{grodsky}, because dual representations are finite-dimensional.
However having a physically acceptable spectrum this equation admits unitarity
only at zero order of $v/c$ \cite{sav13}. In the article \cite{sav14} we found 
a new equation which corresponds to a symmetric solution of the Majorana 
commutation relations and admits unitarity in all orders of $v/c$.
It is based on the same dual representations $(1/2, \pm (1/2 + r))$ of the 
Lorentz algebra and is a natural extension of the previous equation \cite{sav13}
because the new gamma matrices have the same form as the old ones 
but with additional antidiagonal elements. 
The  new gamma matrices are anticommuting  and satisfy generalized 
Clifford algebra \cite{sav14}
\be
 \{ \Gamma_{\mu}, \Gamma_{\nu} \}  = 2~\eta_{\mu\nu} \Gamma^{2}_{0}. \label{atic}
\qee
They guarantee unitarity of the theory at all orders of $v/c$. 
The anticommutation relations (\ref{atic}) are covariant, because the matrices 
$\Gamma^{'}_{\mu} = \Lambda_{\mu}^{~\nu}~ \Gamma_{\nu}$ satisfy the same 
relations (\ref{atic}) as one can check by direct computation (notice that
$d \cdot \Gamma^{2}_{0} = \sum_{\mu} \Gamma^{2}_{\nu}$).
For the completeness we shall review the logical and analytical steps which 
lead to equation presented in \cite{sav13} and then derive 
the new equation which has anticommuting gamma matrices \cite{sav14}.

\subsubsection{Lorentz algebra.}In terms of $SO(3)$ 
generators~ $\vec{a}$~ and Lorentz 
boosts~ $\vec{b}$~~~($a_{x}
= iI^{23}~~$$a_{y} = iI^{31}~~~a_{z} = iI^{12}~~~b_{x} 
 = iI^{10} ~~~ b_{y}=  iI^{20}~~~b_{z} = iI^{30}$)~~the
algebra of the $SO(3,1)$ generators 
can be rewritten as \cite{majorana,heisenberg,dirac} 
(we use Majorana's notations)
\be
[a_{x},  a_{y}]= ia_{z}~~~~[a_{x},  b_{y}]= ib_{z}~~~~[b_{x},  b_{y}]= 
-ia_{z} . \label{alge3} 
\qee
The irreducible representations $R^{(j)}$ of the $SO(3)$ subalgebra 
(\ref{alge3}) are well known. The dimension of  $R^{(j)}$ is~~ $2j+1$~~ and 
$j =0,~ 1/2,~1,~3/2,~...$ The representation $\Theta = (j_{0};\lambda)$ 
of the Lorentz algebra can be parameterized as the sum 
\cite{heisenberg,majorana,dirac}
$
\Theta(j_{0},\lambda) = \oplus \sum^{\infty}_{j=j_{0}}~R^{(j)},$
where $j_{0}$ defines the lower spin in the representation 
and $\lambda$ is a free complex parameter. The
$\lambda$ appears as an essential dynamical parameter which 
cannot be determined solely from the kinematics of the Lorentz group
\footnote{The representation is finite-dimensional if $\lambda=j_{0} +r$, 
$r=1,2,3,...$. The representations used in the 
Dirac equation are $(1/2,-3/2)$ and $(1/2,3/2)$ and
in the Majorana equation they are $(0,1/2)$ in the boson case and $(1/2,0)$ in
the fermion case. The infinite-dimensional Majorana representation $(1/2,0)$ 
contains $j=1/2,3/2,...$ 
multiplets of the $SO(3)$ while $(0,1/2)$ contains $j=0,1,2,...$ multiplets.}. 
The adjoint representation is defined as
$\dot{\Theta} = (j_{0};-\lambda)$. We shall consider the case 
$\Theta_{r} =(r-1/2,\lambda)$   and  
$-3/2 \leq \lambda \leq 3/2 $ to have real matrix elements of the operator
$b_{k}, k=x,y,z$ \cite{sav13}. 

\subsubsection{Superposition principle and physical constraints.}The Majorana 
commutation relation (\ref{mcr}) 
together with the last equations allow to find 
$\Gamma_{\mu}$ matrices when a representation $\Theta$ of the Lorentz algebra 
$I_{\mu\nu}$ is given \cite{majorana}. 
Because $\Gamma_{0}$ commutes with spatial rotations $\vec{a}$ (see (\ref{mcr}))
it should have the form
\be
<j,m\vert~\Gamma^{rr'}_{0}~\vert j'm'> = \gamma^{r r'}_{j} \cdot
\delta_{j j'}\cdot \delta_{m m'}~~~~~~~r,r' = \dot{N},...,\dot{1},
1,...,N   \label{solution1}
\qee
where we consider $N$ pairs of adjoint representations 
$\Theta=(\Theta_{\dot{N}},\cdots,
\Theta_{\dot{1}},\Theta_{1},\cdots, \Theta_{N})$ with $j_{0}= 
1/2,...,N-1/2$.
Thus $\gamma^{rr'}$ is $2N \times 2N$ matrix which 
should satisfy the equation for $\Gamma_{0}$ (\ref{mcr})\cite{majorana}
\be
\Gamma_{0}b^{2}_{z}~ -~ 2~b_{z}\Gamma_{0}b_{z}~  + ~b^{2}_{z}\Gamma_{0}~ = ~-
\Gamma_{0}. \label{funeq}
\qee 
If $\Gamma^{(1)}_{\mu}$ and $\Gamma^{(2)}_{\mu}$ are two
solutions of the equation (\ref{funeq}), then their sum is also a solution 
of (\ref{funeq}) ~\cite{sav13} and we shall use this superposition property 
in order to find solution with required physical properties. 
In \cite{sav13} the authors were searching the solution
of the above equation in the form of Jacoby matrices  
\be
\gamma_{j} = \left( \begin{array}{c}~~~0~~~~,~~~~~\gamma_{j}^{NN-1}
,~~~~~~~~~~~~~~~~~~~~~~~~~~~~~~~~~~~~~~\\
\gamma_{j}^{N-1N},~~~0~~~,~~~\gamma_{j}^{N-1N-2}
,~~~~~~~~~~~~~~~~~~~~~~~~~~~~~\\................................
..................\\..................................
\\~~~~~~~~~~~~~~~~~~~~~~~~~~~~\gamma_{j}^{\dot{N-1}\dot{N-2}}
,~~~0~~~,~~~~~\gamma_{j}^{\dot{N-1}\dot{N}}
\\~~~~~~~~~~~~~~~~~~~~~~~~~~~~~~~~~~~~~~~\gamma_{j}^{\dot{N}\dot{N-1}}
,~~~~0~~~~~\end{array} \right),~~~\Psi_{j}=\left( \begin{array}{c}
        \psi^{N}_{j}\\
        \psi^{N-1}_{j}\\
         .....\\
         .....\\
        \psi^{\dot{N}}_{j}\\
        \psi^{\dot{N-1}}_{j}
\end{array} \right). \label{ansatz}
\qee
In the subsequent work \cite{sav14} we found the finale solution which 
has additional 
nonvanishing antidiagonal elements $\gamma^{r\dot{r}}_{j}$ and represents 
a symmetric solution of the equation (\ref{funeq}) of non-Jacobian form. 

These solutions of the equation (\ref{funeq}) are defined up 
to a set of constant factors which are independent from $j$.
Indeed, because Jacoby matrices (\ref{ansatz}) have a 
specific form, the original equation (\ref{funeq}) factorizes into separate 
equations for every element $\gamma^{rr+1}_{j}$ of the Jacoby matrix 
and the solution has the form \cite{sav13}
\be
\gamma^{rr+1}_{j}~=~Const~\sqrt{(1-\frac{r^{2}}{N^{2}})}~\cdot
\sqrt{(\frac{j^2 +j}{4r^2 -1} -\frac{1}{4})}. \label{jacsolu}
\qee
It has  $(4N-2)$ {\it j-independent free constant}
(this fact reflects the superposition property of the equation 
(\ref{funeq})). This 
freedom allows us to impose necessary physical constraints on a solution 
requiring \cite{sav13,sav14}: 
\begin{emphit}
\item  i)~~physical behaviour of the spectrum, 
\item  ii)~~Hermitean property of the system, 
\item  iii)~~reality and positivity of the probabiity density matrix 
\end{emphit}
This means that we require sensible physical
behaviour of the theory following one-particle interpretation of the solutions
\cite{dirac,pauli}. 

In order to impose  these constraints one should study the 
spectral properties of the matrices of infinite size with matrix elements 
$\gamma^{rr+1}_{j}$ and $\gamma^{r\dot{r}}_{j}$, which have complicated "root" 
dependence. The first inspection of the solution (\ref{jacsolu}) 
simply shows that every element  
$\gamma^{rr+1}_{j}$ grows like $\approx j$ and in general  all eigenvalues 
$\epsilon_{j}$ will also grow with j. 
Therefore the mass spectrum $ M_{j} =  M/\epsilon_{j}$ will have Majorana-like
behaviour (6) $ M_{j} \approx  M/j $. To avoid this general behaviour 
of the spectrum one should carefully inspect eigenvalues of the matrix 
$\Gamma_{0}$ for small values of $N$ and then for arbitrary N \cite{sav13}. The 
parameter N plays the role of a natural regularization.
The $B-H-\Sigma-\Sigma_{1}-\Sigma_{2}$-solutions which appear 
(see \cite{sav13} and below) have exeptional 
behaviour: half of the eigenvalues of the spectrum are increasing and 
the other half are decreasing. One can achieve this exeptional 
behaviour of the solution by tuning the free constants in the  
general solution (\ref{jacsolu}). However  solutions 
$B-H-\Sigma-\Sigma_{1}-\Sigma_{2}$ cannot be 
accepted \cite{sav13} because still half of the eigenvalues produce a mass 
spectrum which has an accumulation point at zero mass. 

\subsubsection{Dual representations.} Last phenomenon can 
be understood on the example of the Dirac equation. For that let us define
the {\it dual representation} as \cite{sav13}
$\Theta = (j_{0};\lambda)  \rightarrow (\lambda;j_{0})
= \Theta^{dual}$. This symmetry transformation imposes constraints on the 
free parameter $\lambda$, so that it should be {\it integer or half-integer}.
This is because in the representation $(j_{0};\lambda)$~~ $j_{0}$ must be 
integer or half-integer \cite{dirac1}. For the dual representations $\Theta$ 
and $\Theta^{dual}$ the matrix elements of Lorentz
generators $I_{\mu\nu}$ are  precisely the same, the only difference 
between them is that the lower spin is equal to $j_{0}$ for the reperesentation 
$\Theta$ and is equal to $\lambda$ for its dual $\Theta^{dual}$. 
Therefore any solution $\Gamma_{\mu}$
of the Majorana commutation relations (\ref{mcr}) for  $\Theta$ can be 
translated into the corresponding solution $\Gamma^{dual}_{\mu}$
for $\Theta^{dual}$ by exchanging $j_{0}$ for $\lambda$ \cite{sav13}.

The dual transformation
of the Dirac representations  $(1/2,-3/2)$ and $(1/2,3/2)$ would be 
infinite-dimensional  $(3/2,-1/2)$ and $(3/2,1/2)$ with 
$j=3/2,5/2,...$ and the corresponding solution $\Gamma^{dual}_{0}$ 
has the form 
$\gamma^{1~\dot{1}}_{j} = \gamma^{\dot{1}~1}_{j} = j+1/2$ with the 
following mass spectrum 
\be
M^{Dirac~dual}_{j}=\frac{M}{j+1/2},~~~~~~~~ j=3/2,5/2,....    \label{10}     
\qee
This Majorana-like mass spectrum is dual to the physical spectrum 
of the Dirac equation
\be
M^{Dirac}_{j}= M,~~~~~~~~ j=1/2.       \label{11}  
\qee
The dual equation is simply unphysical, but we have to admit   
that the whole decreasing mass spectrum of the dual equation 
{\it corresponds or is dual} to a physical Dirac fermion. 
From this point of view we have to ask  about 
physical properties of the equations which are dual to "unphysical" ones 
$B-H-\Sigma-\Sigma_{1}-\Sigma_{2}$. The dual 
transformation completely improves the decreasing mass spectrum 
of these equations \cite{sav13}, as it takes place  in (\ref{10}) and (\ref{11}),
and indeed the last $\Sigma^{dual}_{2}$-equation has the spectrum of particles and
antiparticles of increasing half-integer spin lying on quasilinear trajectories
of different slope. However having a physically acceptable spectrum 
(see constraint i)~) this 
equation admits  unitarity (see constraint ii)~) only at zero order of $v/c$. The  
equation which admits unitarity at all orders have been found in \cite{sav14} 
and we shall review it below.

\subsubsection{Invariant product and probability density.} Let us introduce 
the invariant scalar product~$
<~\Theta ~\Psi_1~\vert~ \Theta ~\Psi_2~>~=~<~\Psi_1~\vert~ \Psi_2~> 
$, where $\Theta = 1 + \frac{1}{2}\epsilon_{\mu \nu}~I^{\mu \nu}$ and  
the matrix $\Omega$ is  defined as
$
<~\Psi_1~\vert~ \Psi_2~> = \bar{\Psi}_{1}~\Psi_{2} =\Psi^{+}_{1}~\Omega~\Psi_{2}
 = \Psi^{*~r}_{1~jm}~\Omega^{rr'}_{jm~j'm'}~\Psi^{r'}_{2~j'm'}
$ with the properties
\be
\Omega~a_{k} = a_{k}~\Omega~~~~\Omega~b_{k} = b^{+}_{k}~\Omega ~~~~\Omega 
= \Omega^{+}.     \label{gode} 
\qee
From the first relation it follows that 
$\Omega = \omega^{rr'}_{j}\cdot\delta_{jj'}\cdot\delta_{mm'}$
and from the last two equations, for our choice of the representation $\Theta$
and for a real $\lambda$ in the interval $-3/2 \leq \lambda \leq 3/2 $, that
$
\omega^{r\dot{r}}_{j}=\omega^{\dot{r}r}_{j}~=~1~~~\omega^{2}_{j}=1,
$ thus $\Omega$ is an antidiagonal matrix. The conserved current density 
is equal to
$J_{\mu} = \bar{\Psi}~\Gamma_{\mu}~\Psi,~~\partial^{\mu}~J_{\mu}=0.$
The current density $J_{0}$ should be {\it real and positive definite} 
(see constraint iii)~), which is equivalent to the requirement that 
\be
\Gamma^{+}_{\mu}~\Omega =\Omega~\Gamma_{\mu}, \label{holdc}
\qee
and to the positivity of the eigenvalues of the matrix 
$\rho = \Omega~\Gamma_{0}$.

\subsubsection{The $B-H-\Sigma-\Sigma_{1}-\Sigma_{2}$-solutions.}
The basic solution ( {\it B-solution}) of the equation (\ref{funeq}) for the 
$\Gamma_{0}$ has the form (\ref{ansatz}), (\ref{jacsolu}) with all set of constant 
factors equal to $Const=i$ \cite{sav13} and 
$
\gamma^{1~\dot{1}}_{j} = \gamma^{\dot{1}~1}_{j} = j+1/2
$.
The positive eigenvalues $\epsilon_{j}$ can be found in \cite{sav13}.
They  show that the coefficient of 
proportionality behind $j$ drops N times compared with the one in 
the Majorana solution $\epsilon_{j}=j+1/2$ in (6)
and now  many eigenvalues are less 
than unity and the corresponding masses $M_{j}= M/\epsilon_{j}$
are bigger than the ground state mass M. This actually means 
that increasing the number of representations in $\Theta=
(\Theta_{\dot{N}},\cdots,\Theta_{\dot{1}},\Theta_{1},\cdots, \Theta_{N})$ 
one can slow down the growth of the eigenvalues. To have  
the mass spectrum  bounded from below one should have  spectrum with 
all eigenvalues $\epsilon_{j}$ less than unity.

In the limit $N \rightarrow \infty$ the B-solution 
is being reduced to the form
\be
\gamma^{r+1~r}_{j}=\gamma^{r~r+1}_{j} =
\gamma^{\dot{r+1}~\dot{r}}_{j}= \gamma^{\dot{r}~\dot{r+1}}_{j} = 
i~ \sqrt{(\frac{j^2 + j}
{4r^2-1} -\frac{1}{4})}~~~~~j\geq r+1/2  
\qee
and
$
\gamma^{1~\dot{1}}_{j} = \gamma^{\dot{1}~1}_{j} = j+1/2
$,
where $r=1,2,...$. All eigenvalues $\epsilon_{j}$ 
tend to unity when the number of representations $N \rightarrow \infty$.
The characteristic equation which is satisfied by the gamma matrix in 
this limit is
\be
(\gamma_{j}^2 -1)^{j+1/2} = 0 ~~~~~~~~~~j=1/2,~3/2,~,5/2,\cdots \label{det1}
\qee
with all eigenvalues $\epsilon_{j} = \pm 1$. Therefore all states have equal
masses $M_{j}= 1$, but the Hamiltonian 
is not Hermitian ($\Gamma^{+}_{0} \neq \Gamma_{0}$) \footnote{The determinant
and the trace are equal to
$
Det~\gamma_{j} = \pm 1,~~Tr~\gamma^{2}_{j} = 2j +1,  
$
thus~$
\epsilon_{1}^2 \cdot ...\cdot\epsilon_{j+1/2}^2= 1,~~~\epsilon_{1}^2 + 
...+\epsilon_{j+1/2}^2 = j+1/2.
$}.
The matrix $\Omega~\Gamma_{0}$ has the characteristic equation
$
(\omega_{j}~\gamma_{j} - 1)^{2j + 1} = 0
$
with all eigenvalues equal to $\rho_{j} = +1$.
Thus the matrix $\Omega~\Gamma_{0}$ is positive definite and all its 
eigenvalues are equal to one, but the relations
$
\Omega~\Gamma_{0} \neq \Gamma^{+}_{0}~\Omega,~~\Gamma^{+}_{0} \neq
\Gamma_{0}
$
do not hold. What is crucial
here is that we can improve the B-solution without 
disturbing its determinant which is equal to one (\ref{det1}) 
$(Det\Gamma_{0} =1)$.  The last property of the determinat is  
necessary to keep, in order that the spectrum will be symmertically 
distributed arround unity. 

The Hermitian solution ({\it H-solution}) of (\ref{funeq}) for $\Gamma_{0}$
can be found as a phase modification of the basic {\it B-solution} \cite{sav13}
\be
\gamma^{r+1~r}_{j}=-\gamma^{r~r+1}_{j} =
-\gamma^{\dot{r+1}~\dot{r}}_{j}= \gamma^{\dot{r}~\dot{r+1}}_{j} =
i~ \sqrt{(\frac{j^2 + j}{4r^2-1} -\frac{1}{4})}~~~~~j\geq r+1/2 .  \label{herso}
\qee
These matrices are Hermitian $\Gamma^{+}_{0}=\Gamma_{0}$, but
the characteristic equations are more complicated now. 
These  polynomials $p(\epsilon)$ have reflective symmetry and are even
$p_{j}(\epsilon)~=$ $\epsilon^{2j+1}$~ $p_{j}(1/\epsilon)$, 
$p_{j}(-\epsilon)~= ~p_{j}(\epsilon)$ therefore 
if $\epsilon_{j}$ is a solution then $1/\epsilon_{j}$, ~$-\epsilon_{j}$
and $-1/\epsilon_{j}$ are also solutions \footnote{
Computing the traces and determinants of these matrices one can get 
the following general relation for the eigenvalues
$
\epsilon_{1}^2 \cdot ...\cdot\epsilon_{j+1/2}^2= 1,~~~\epsilon_{1}^2 + 
...+\epsilon_{j+1/2}^2 = j(2j+1).
$}.
The eigenvalues $\epsilon_{j}$ can be found in \cite{sav13}.
The changes in the phases of the matrix elements (\ref{herso}) 
result in different behaviour of eigenvalues.
The half of the eigenvalues (decreasing eigenvalues) produce quasilinear 
trajectories with nonzero string tension
and the other half (increasing eigenvalues) affect the low spin states 
on  trajectories, so that the smallest mass on a given trajectory  
tends to zero (see \cite{sav13}). The matrix $\Omega~\Gamma_{0}$ has again the 
characteristic equation
$
(\omega_{j}~\gamma_{j} - 1)^{2j + 1} = 0
$
and all eigenvalues are equal to one. Thus again the  
matrix $\Omega~\Gamma_{0}$ is positive definite 
because all eigenvalues are equal to one, but the important relation
$\Omega~\Gamma_{0} \neq \Gamma^{+}_{0}~\Omega$
does not hold. 

The  solution of (\ref{funeq}) for $\Gamma_{0}$ with both properties
$\Gamma^{+}_{0} = \Gamma_{0}$ and $\Omega~\Gamma_{0} = \Gamma^{+}_{0}~\Omega$
can be found by using the basic solution rewritten with
arbitrary phases of the matrix elements and then by requiring that $\Gamma_{0}$
should be Hermitian $\Gamma^{+}_{0} = \Gamma_{0}$ and should satisfy the 
relation $\Omega~\Gamma_{0} = \Gamma^{+}_{0}~\Omega$. This solution,  
$\Sigma$-$solution$, is symmetric and has the form \cite{sav13}
\be
\gamma^{r+1~r}_{j}=~\gamma^{r~r+1}_{j} =
\gamma^{\dot{r+1}~\dot{r}}_{j}= \gamma^{\dot{r}~\dot{r+1}}_{j} =
 \sqrt{(\frac{j^2 + j}{4r^2-1} -\frac{1}{4})}~~~~~j\geq r+1/2.   \label{realsym}
\qee
In this case the Hermitian matrix $\Gamma^{+}_{0} = \Gamma_{0}$ 
has the desired property 
$
\Gamma^{+}_{0}~\Omega = \Omega~\Gamma_{0}.
$
This  means that the current density is equal to $\rho =\Omega~\Gamma_{0}$.
In addition, all of the gamma matrices now have this property (\ref{holdc})
$
\Gamma^{+}_{k}~\Omega = \Omega~\Gamma_{k}~~k=x,y,z
$
which follows from the equation $\Gamma_{k}= i [b_{k}~\Gamma_{0}]$ (\ref{mcr}) 
and equation (\ref{gode}) $\Omega~b_{k}~ = b^{+}_{k}~\Omega$.

The characteristic equations 
and the spectrum are the same for the Hermitian 
H-solution and symmetric $\Sigma$-solution, but 
the corresponding characteristic equations for the matrices $\rho_{j}$ are
different and the eigenvalues are not positive
definite any more \cite{sav13}, that is there are many ghost states.
The positive norm physical states are lying on 
the quasilinear trajectories of different slope and  
the negative norm ghost states are also lying on the quasilinear 
trajectories. There are also tachyonic solutions \cite{sav13},
which appear in Majorana equation as well \cite{majorana,dirac1,grodsky,sav13}. 
In \cite{sav13} it was suggested to "protect" equation from tachyonic solutions
by setting some of the transition amplitudes $\gamma^{r~r'}_{j}$ to zero.

In the case when some of the transition amplitudes $\gamma^{r~r'}_{j}$ 
in (\ref{realsym}) are set to zero (we again use "superposition" property 
of the equation (\ref{funeq}) \cite{sav13})
\be
\gamma^{\dot{1}~1}~ = ~ \gamma^{2~3}_{j}~=~\gamma^{4~5}_{j} 
= ....=0~~~~~\gamma^{1~\dot{1}}~=~\gamma^
{\dot{2}~\dot{3}}~=~\gamma^{\dot{4}~\dot{5}}~= ... = 0 \label{diaggammsq}
\qee
and all other elements of the $\Gamma_{0}$ matrix remain the same as in 
(\ref{realsym}) we have a new $\Sigma_{1}$-solution with the important 
property that $\Gamma^{2}_{0}$ is a diagonal matrix and that the antihermitian 
part of $\Gamma_{k}$ anticommutes with $\Gamma_{0}$. Thus in
this case we recover the nondiagonal 
part of the Dirac commutation relations for gamma matrices
$
\{ \Gamma_{0}, \tilde{\Gamma_{k}} \} =0~~~~~~~~k =x,y,z. 
$
For the solution (\ref{diaggammsq}) 
one can explicitly compute the mass spectrum and the slope of the trajectories 
\cite{sav13}
\be
M^{2}_{n}= \frac{2 M^{2}}{n}~ \frac{j^2 -(2n-1)j 
+n(n-1)}{j-(n-1)/2}~~n=1,2,... \label{strtension}
\qee
where $j = n+1/2, n+5/2, ....$.
The string tension $\sigma_{n} = 1/2\pi\alpha^{'}_{n}$
varies from one trajectory to another and is equal to
\be
2\pi \sigma_{n}  = \frac{1}{\alpha^{'}_{n}} = \frac{2M^2}{n}~~~~~~n=1,2,... 
\label{s}
\qee
Thus we have the string equation which has 
trajectories with different string tension and that trajectories 
with large $n$
are almost "free" because the string tension tends to zero.
The smallest mass   
on a given trajectory $n$ has spin $j = n+1/2$ and decreases as 
$
M^{2}_{n}(j=n+1/2)= \frac{3M^2}{n(n+3)}. 
$ 

The other solution, $\Sigma_{2}$-solution, which shares the above properties 
of $\Sigma_{1}$-solution is (\ref{realsym}) with 
\be
\gamma^{1~2}_{j}~ = ~ \gamma^{3~4}_{j}~= ....
=0~~~~~~\gamma^{\dot{1}~\dot{2}}_{j}~=~\gamma^{\dot{3}~\dot{4}}_{j}~= ... = 0.
\label{delta2}
\qee
The difference between these last two solutions is that in the first case 
the lower spin is $j=3/2$ and in the second case it is $j=1/2$.
The unwanted property of all these solutions $\Sigma$, $\Sigma_{1}$ and 
$\Sigma_{2}$ is that the  smallest mass 
$M^{2}_{n}(min)$ tends to zero and the spectrum still has an accumulation point.
We have to remark also that both equations, $\Sigma_{1}$ and $\Sigma_{2}$, 
which correspond to  (\ref{diaggammsq}) 
and to (\ref{delta2}) have natural $constraints$ \cite{sav13}. 

\subsubsection{Dual equations.} The unwanted property of the $\Sigma$-solutions, 
that is the  decreasing of the smallest mass
on a given trajectory, can be avoided by dual transformation of the system 
\cite{sav13} (see section "Dual representations"). Under the dual 
transformation  
$
\Theta = (j_{0};\lambda)  \rightarrow (\lambda;j_{0})
= \Theta^{dual}
$
the representation $\Theta=(\Theta_{\dot{N}},\cdots,
\Theta_{\dot{1}},\Theta_{1},\cdots, \Theta_{N})$  is transformed into its dual
$\Theta^{dual}=
.....(\lambda; -5/2)~~$ $(\lambda; -3/2)~~$ $(\lambda; -1/2)~~(\lambda; 
1/2)~~$ $(\lambda;3/2)~~(\lambda; 5/2)....$ 
and we are lead to take $\lambda$ to be half-integer and to 
$\lambda = 1/2$ in order to have the Dirac
representation incorporated in $\Theta$ \footnote{These representations do 
not coinside with the ones in Ramond equation \cite{ramond}.}. 
The solution which is dual to 
$\Sigma_{2}$
(\ref{realsym}) and  (\ref{delta2})  is equal to \cite{sav13}
\be
\gamma^{r+1~r}_{j}=~\gamma^{r~r+1}_{j} =
\gamma^{\dot{r+1}~\dot{r}}_{j}= \gamma^{\dot{r}~\dot{r+1}}_{j} =
 \sqrt{(\frac{1}{4}-\frac{j^2 + j}{4r^2-1})}~~~~~r\geq j+3/2    \label{dualsym}
\qee
where $j=1/2,3/2,5/2,...$,~~$r=2,4,6,....$ and the rest of the elements are equal to zero
\be
\gamma^{\dot{1}~1}~ = ~ \gamma^{1~2}_{j}~=~\gamma^{3~4}_{j} 
= ....=0~~~~~\gamma^{1~\dot{1}
}~=~\gamma^{\dot{1}~\dot{2}}~=~\gamma^{\dot{3}~\dot{4}}~= 
... = 0.\label{dualsym1}
\qee
The Lorentz boost operators $\vec{b}$ are antihermitian in this case 
$b^{+}_{k} = -b_{k}$ \cite{sav13}, and 
therefore the $\Gamma_{k}$ matrices are also antihermitian
$\Gamma^{+}_{k} = - \Gamma_{k}$.
The matrix $\Omega$ changes and is now equal to the parity operator $P$, the
relation $\Omega~\Gamma^{+}_{\mu}= \Gamma_{\mu}~\Omega$
remains valid. The diagonal part of $\Gamma_{k}$ anticommutes
with $\Gamma_{0}$
as it was before $\{ \Gamma_{0}, \tilde{\Gamma_{k}} \} =0~~k =x,y,z$.
The mass spectrum is equal to
\be
M^{2}_{n}= \frac{2M^2}{n}~\frac{(j+n)(j+n+1)}{j+(n+1)/2} 
\label{dualmassspectrum}
\qee
where $n=1,2,3,..$ and enumerates the trajectories. 
The lowest spin on a given trajectory is either $1/2$ or $3/2$ 
depending on n: if n is odd then 
$j_{min}=1/2$, if n is even then $j_{min}=3/2$. This is an essential
new property
of the dual equation because now we have an infinite number of states 
with a given
spin $j$ instead of $j+1/2$.
The string tension is the same as in the dual system  (\ref{s}).  
The lower mass on a given trajectory $n$ is given by the formula $(j=1/2)$ is
$
M^{2}_{n}(j=1/2) = \frac{4M^2}{n}\frac{(2n+1)(2n+3)}{n+2} \rightarrow (4M)^2
$ 
and the spectrum is bounded from below by positive mass.

\subsubsection{Generalized wave equation.} The last $\Sigma^{dual}_{2}$-equation 
has the property that only the diagonal matrix elements of the anticommutator
$\{ \Gamma_{0}, \Gamma_{z} \}$ are equal to zero ~$
<j,m,r \vert \{ \Gamma_{0}, \Gamma_{z} \} \vert r,j,m> =0,
$
and that nondiagonal elements are not equal to zero  
$
<j-1,m,r \vert \{ \Gamma_{0}, \Gamma_{z} \} \vert r,j,m> = 
i\sqrt{j^2 - m^2}~\varsigma^{r}_{j}~[ (\gamma^{rr+1}_{j})^{2} - 
(\gamma^{rr+1}_{j-1})^{2}].
$
Let us search the solution of the Majorana commutation relation (\ref{funeq})
in the same $\Sigma^{dual}_{2}$-Jacoby form (\ref{ansatz}) but with 
additional nonvanishing antidiagonal matrix elements $\gamma^{r~\dot{r}}_{j}$. 
The solution has the form \cite{sav14}
\be
\gamma^{r~\dot{r}}_{j}=~\gamma^{\dot{r}~r}_{j}=~-\gamma^{r+1~\dot{r+1}}_{j} 
=~-\gamma^{\dot{r+1}~r+1}_{j}= \frac{j + 1/2}
{\sqrt{4r^2-1}}
\qee
where $j=1/2,3/2,5/2,...$,~~$r=2,4,6,....$ and $r \geq j +3/2$ and one can check
directly that $\Gamma_{0}$ is the solution of (\ref{funeq}). These additional 
matrix elements in $\Gamma_{0}$ will not change the diagonal matrix elements 
of the anticommutator but  will cancel nondiagonal matrix elements \cite{sav14}
$
<j-1,m,r \vert \{ \Gamma_{0}, \Gamma_{z} \} \vert r,j,m> =0. 
$
One can check this fact also using the relation 
$\{ \Gamma_{0}~\Gamma_{z} \}=i~[b_{z}~\Gamma^{2}_{0}]$, which follows from
(\ref{mcr}).
Using the relations (\ref{mcr}) $\Gamma_{y} =-i~[\Gamma_{z}~a_{x}]$ and 
$[\Gamma_{0}~a_{x}]=0$ one can see that $\{ \Gamma_{0}, \Gamma_{y} \}=0$ and 
in the same way that $\{ \Gamma_{0}, \Gamma_{x} \}=0$. Finally using the 
relation (\ref{mcr}) $\Gamma_{k} =-i~[\Gamma_{0}~b_{k}]$ one can prove by 
direct calculation
that  $\{ \Gamma_{k}, \Gamma_{l} \}=0$ for $k\neq l$ and then using 
(\ref{funeq}) and the fact that $[b_{k}~\Gamma^{2}_{0}]=0$  one can prove that 
$\Gamma^{2}_{k}=-\Gamma^{2}_{0}$, therefore \cite{sav14}
\be
\{ \Gamma_{\mu}, \Gamma_{\nu} \}~=~2 \eta_{\mu\nu}~ \Gamma^{2}_{0}, 
\qee
where $\Gamma^{2}_{0}$ is a diagonal martix. Now the theory is 
Hermitian in all orders of $v/c$.
The mass spectrum is highly degenerated and is given by the formula
\be
M^{2}_{j}= M^{2} (1~-~\frac{1} {4r^{2}} )~~~~~~~~r \geq j +1/2.
\qee
New  mass terms $(\vec{a}\cdot\vec{b})~\Gamma_{5}$ and $(\vec{a}^2 -\vec{b}^2)$ 
can be added into the string equation (\ref{stringeq})
in order to increase the  string tension 
\be
\{~i~\Gamma_{\mu}~\partial^{\mu}~~-~~M~(\vec{a}
\cdot\vec{b})\Gamma_{5}~~-~~gM~(\vec{a}^2 -\vec{b}^2)~\}~~\Psi~~~=0,  
\qee
where $(\vec{a}\cdot\vec{b})$ and $(\vec{a}^2 -\vec{b}^2)$ are the Casimir 
operators of the Lorentz algebra. Now all trajectories acquire a nonzero slope
\be
M^{2}_{j}= M^{2}~\frac{4r^{2}-1}{4r^{2}} (r-1/2)^{2}
(1+ 2g (r-1/2))^{2}~~~~~~~r\geq j+1/2.
\qee
where $ r=2,4,6,....$,~~$j=1/2,3/2,5/2,....$,
thus $M^{2}_{j} \geq M^{2}(j+1)^2$.
Thus the  spectrum of the theory consists of 
particles and antiparticles of increasing half-integer spin 
and masses.
On every mass level there are particles with spin running from 
$j=1/2$ up to $j_{max}=r-1/2$, $r=2,4,..$ and the degeneracy is equal 
to $d_{r}=2r-1 = 2j_{max}$. This is in contrast with the exponential 
degeneracy in case of superstrings. 
The tachyonic solutions which appear in Majorana equation (see (20) in 
\cite{majorana}) do not show up here. 

We can introduce now the interaction with gauge field using 
covariant derivative $\Pi_{\mu} = i \partial_{\mu} + g A_{\mu} $ 
and see that there are no obvious inconsistencies which are characteristic 
to high spin equations in the background 
field (see W.Pauli and M.Fierz in \cite{dirac1}), 
because the eqaution (\ref{stringeq}) can be transformed to the form
\be
\{~\Pi^{2} ~\Gamma^{2}_{0}~~+~~\frac{ig}
{2}~\Gamma_{\mu}~\Gamma_{\nu}~F^{\mu \nu}~~-~~M^{2}~\}~~\Psi~~~=0  
\qee
and does not produce additional constraints.

In conclusion I would like to acknowledge Holger Nielsen and Konstantin Savvidy 
for stimulating discussions. I wish to thank Costas Kounnas and  George Koutsoumbas 
for support. I would like to acknowledge the organizers of the conference for 
inviting me and for arranging an interesting meeting. 
%
%

\end{document}